%% file: main.tex
\documentclass{article} 
\usepackage{iclr2023_conference,times}

\input{math_commands.tex}

\usepackage{hyperref}
\usepackage{url}

\usepackage[utf8]{inputenc} 
\usepackage[T1]{fontenc}    
\usepackage{amsfonts}       
\usepackage{amsmath}
\usepackage{amssymb}
\usepackage{nicefrac}       
\usepackage{microtype}      
\usepackage{bm}
\usepackage[normalem]{ulem}
\usepackage{mathtools}
\usepackage{color}
\usepackage{array}
\usepackage{caption}
\usepackage{subcaption}

\usepackage[pdftex]{graphicx}
\usepackage{booktabs} 
\usepackage{makecell}
\usepackage{tabularx}
\usepackage{longtable}
\usepackage{multirow}
\usepackage{wrapfig}

\newcommand{\june}[0]{{\textsc{June} }}

\title{Mitigating Disease Spread by Design \\ in Refugee and IDP Camps}

\author{Giulia Zarpellon$^1$, Joseph Aylett-Bullock$^{1,2}$, Frank Krauss$^2$ \& Miguel Luengo-Oroz$^1$  \\ 
$^1$United Nations Global Pulse, $^2$Institute of Data Science, Durham University\\
\texttt{giulia@unglobalpulse.org}, \texttt{j.aylettbullock@bristol.ac.uk}, \\ \texttt{frank.krauss@durham.ac.uk}, \texttt{miguel@unglobalpulse.org} \\
\And
}

\iclrfinalcopy 
\begin{document}

\maketitle

\begin{abstract}
Disease spread represents an increasing challenge in refugee and internally displaced person (IDP) settlements. 
The movement and interaction of people within camps is influenced by their layout, which therefore has the potential to significantly affect disease spread.
This work aims at creating a methodology to explore the potential effects of different camp layouts as mitigating factors in the spread of diseases within settlements. 
We showcase proof-of-concept experiments by leveraging the \june agent-based epidemic model, discuss the kind of operational insights this methodology can facilitate, and provide a framework for future investigations.
\end{abstract}

\section{Introduction}
\label{sec:intro}

Disease spread in refugee and internally displaced person (IDP) settlements represents an increasing challenge for the world’s most vulnerable populations. 
Settlements, especially those rapidly set up in the wake of a crisis, often suffer from overcrowding and insufficient healthcare facilities -- conditions that accelerate disease spread. 
As discussed in~\citep{aylett2022epidemiological}, the likelihood of outbreaks affecting displaced populations is bound to increase in a future with more people far from their homes due to conflict and climate change. 

While access to medication and vaccines remains a key strategy when dealing with infectious diseases~\citep{altare2019infectious}, the COVID-19 epidemic exposed additional challenges that arise when facing a rapidly spreading new disease or in situations without sufficient medical supplies. 
This forces us to think more broadly about mitigation strategies. 
The geography and the organization of a settlement’s layout influences how people move within it and interact with each other; the location of shelters and homes, supermarkets, religious centers, schools, and other structures is a key factor in the spread of infections.  
Developing and testing design paradigms that minimize this spread represents an unique opportunity to embed anticipatory mitigation actions in the construction of new settlements. 




\section{Background}
\label{sec:background}

\textbf{Planning refugee settlements}
Several conceptual frameworks for the design of new camps or the expansion of existing ones have been explored~\citep{jahre2018approaches}. 
In practice, high-level design principles are refined by spatial and geographical factors, constraints to resource allocation and funding, and the critical need to guarantee minimum standards. 
While settlements around the world form in a variety of circumstances, there are general guidelines to site planning and management, such as the Sphere handbook~\citep{sphere} and the Camp Management Toolkit developed by~\citet{unhcr-toolkit}. 
These standards span several aspects of a camp functioning, from recommending a minimum allocation of open space per person, to providing indications on how to best support the population's dietary needs. 

\textbf{Epidemic modeling with \june}
The \june framework~\footnote{Open-sourced at 
\url{https://github.com/idas-durham/june}}~\citep{aylett2021june} originally implemented an agent-based model to study COVID-19 interventions in the UK, modeling a synthetic population with high granularity and simulating individual movements and activities at associated venues with calibrated group-level dynamics. 
We base our work on the adaptation of \june to the case of refugee settlements~\footnote{\url{https://github.com/UNGlobalPulse/UNGP-settlement-modelling}}, developed in collaboration with UN agencies~\citep{aylett2021operational}. 
In the context of the Cox's Bazar refugee settlement in Bangladesh, the model has been calibrated following available research on local and inter-camp mixing patterns~\citep[see][]{walker2022mixed}.

\section{Objectives and methods}
\label{sec:methodology}

The goal of this work is to create a methodology to explore the role of camp layout as a mitigating factor in the spread of infectious diseases within settlements, and showcase proof-of-concept examples of the insights such a framework can generate. 
We rely on the multi-agent \june epidemic modeling framework \citep{aylett2021june} and its adaptation to the case of refugee settlements~\citep{aylett2021operational} to explore prototypical layout options in virtual camps. 

\begin{figure}[t]
\begin{center}
    \begin{subfigure}[b]{0.24\textwidth}
         \centering
         \includegraphics[height=2.6cm]{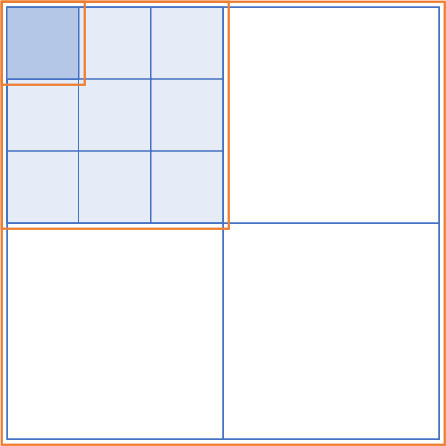}
         \caption{}
         \label{fig:region}
    \end{subfigure}
    \hfill
    \begin{subfigure}[b]{0.24\textwidth}
         \centering
         \includegraphics[height=2.6cm]{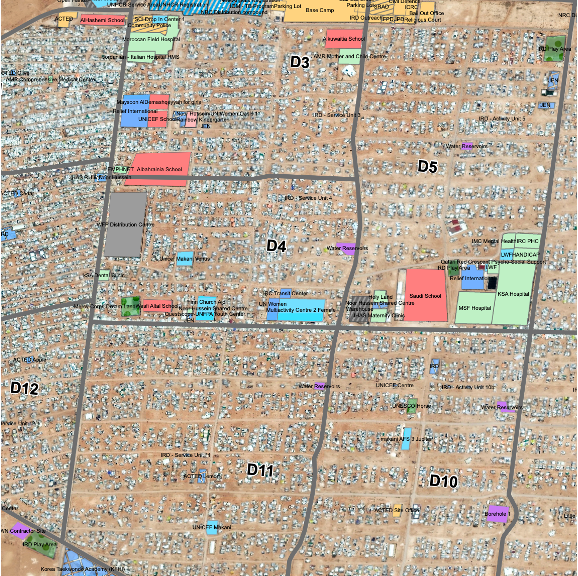}  
         \caption{}
         \label{fig:zaatari-grid}
     \end{subfigure}
     \hfill
     \begin{subfigure}[b]{0.24\textwidth}
         \centering
         \includegraphics[height=2.6cm]{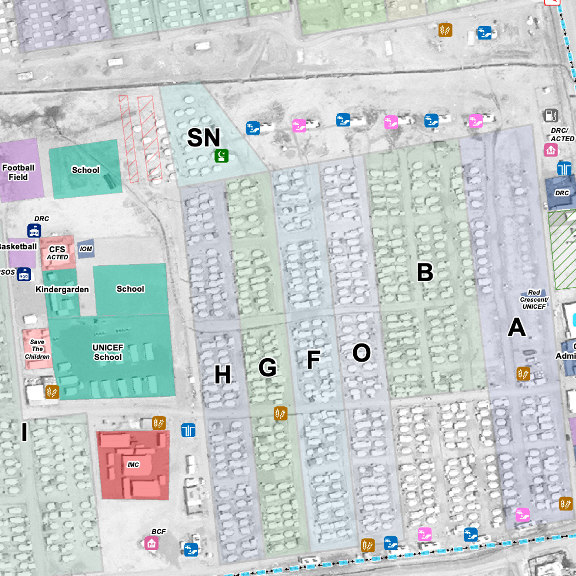}
         \caption{}
         \label{fig:qushtapa-grid}
     \end{subfigure}
     \hfill
     \begin{subfigure}[b]{0.24\textwidth}
         \centering
         \includegraphics[height=2.6cm]{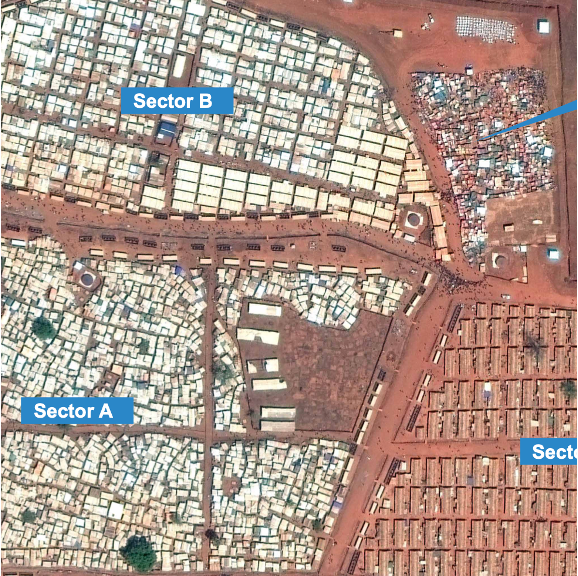}
         \caption{}
         \label{fig:wau-grid}
     \end{subfigure}
\end{center}
\caption{(a) A region building block. 
(b, c, d) Details of grid-like camp structure found in Zaatari (Jordan), Qushtapa (Iraq), Wau (South Sudan). Image details are from \citet{zaatari_grid, qushtapa_grid, wau_grid}. }
\label{fig:grid}
\end{figure}

The framework that we develop to compare different camps layouts comprises four main steps: 


\textbf{\emph{(i)} Defining a geography for the virtual settlement } We define a settlement's geography using square “camp” building blocks. These blocks are intended to represent \emph{regions} of the settlement, and are further subdivided into \emph{super-areas} and \emph{areas}. 
Figure~\ref{fig:region} illustrates these sectors’ hierarchy and grid-like structure. 
Settlements organized in grids are a prototypical module for us to study, and are commonly found around the world (see Figure~\ref{fig:zaatari-grid}, \ref{fig:qushtapa-grid}, \ref{fig:wau-grid}). 

\begin{figure}[t]
\begin{center}
    \begin{subfigure}[b]{0.24\textwidth}
         \centering
         \includegraphics[height=2.6cm]{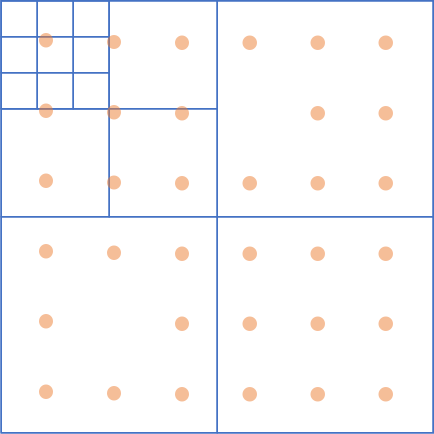}  
         \caption{Even}
         \label{fig:even}
     \end{subfigure}
     \hfill
     \begin{subfigure}[b]{0.24\textwidth}
         \centering
         \includegraphics[height=2.6cm]{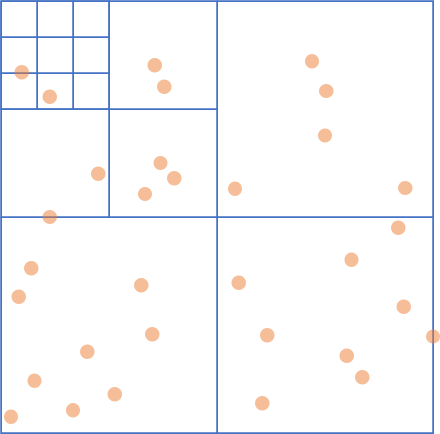}
         \caption{Uniform}
         \label{fig:uniform}
     \end{subfigure}
     \hfill
     \begin{subfigure}[b]{0.24\textwidth}
         \centering
         \includegraphics[height=2.6cm]{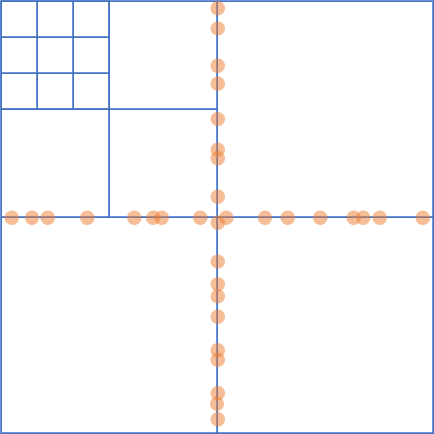}
         \caption{Middle}
         \label{fig:middle}
     \end{subfigure}
     \hfill
     \begin{subfigure}[b]{0.24\textwidth}
         \centering
         \includegraphics[height=2.6cm]{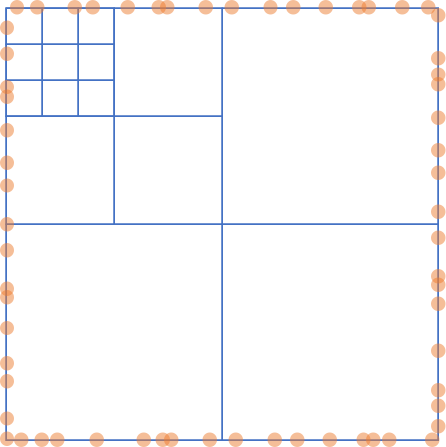}
         \caption{Boundary}
         \label{fig:boundary}
     \end{subfigure}
\end{center}
\caption{Abstractions for venues positioning in square virtual camps with four regions. 
}
\label{fig:positioning}
\end{figure}

\textbf{\emph{(ii)} Generating a synthetic population }
We synthesize population data following the demography of a selected \emph{basecamp}, i.e., a real-world camp for which we know basic population characteristics such as the age/sex distribution and household counts. 
While here we only present experiments with populations sampled from the Zaatari camp (Jordan), we also tested sampling from the Cox’s Bazar (Bangladesh) and Kismayo (Somalia) settlements. 
Appendix~\ref{app:sources} reports details on data sources and the Zaatari basecamp demography. 

\textbf{\emph{(iii)} Distributing shared facilities across the virtual settlement }
During simulations, agents can visit the venues implemented in~\citet{aylett2021operational} and listed in Table~\ref{tab:per-capita} (Appendix~\ref{app:sources}). 
For each facility type, we use a baseline per-capita parameter to determine how many venues should be positioned depending on the synthesized population.  
\newline 
\newline 
We test four archetypes for placing facilities in the virtual geography, illustrated in Figure~\ref{fig:positioning}. 
In the \texttt{even} setting, the prescribed number of venues are placed on a square grid of regularly spaced points within the settlement boundary. 
To place $n$ facilities, we round $n$ to the nearest square number, create a square lattice of locations, and randomly remove extra positions. 
In \texttt{uniform} position, venues locations are uniformly sampled within the camp range. 
Finally, \texttt{middle} and \texttt{boundary} schemes sample locations on the middle axes and the external boundary of the settlement, respectively. 
Note that in the \texttt{uniform}, \texttt{middle} and \texttt{boundary} settings we still place play groups and hand-pumps/latrines \emph{evenly} throughout the camp.

\textbf{\emph{(iv)} Infecting the virtual settlement }
While \june can potentially support different diseases, we test the spread of COVID-19 infections. 
The epidemiological ingredients of the model are extensively discussed in~\citet{aylett2021june,aylett2021operational}. 
Appendix~\ref{app:sources} summarizes our assumptions concerning infection seeding, co-morbidities and interaction matrices. 
To assess the mitigation effects of different camp layouts, we assume \emph{no} containment strategy (e.g., no social distancing).

\section{Experiments}
\label{sec:experiments}
We test virtual settlements composed of 4 or 16 regions arranged in a square, each region extending for 0.81 km\textsuperscript{2} (150 m long areas). 
Assuming population densities compliant with minimum standards~\citep{unhcr-toolkit}, the two settings mainly differ by size of the population and site extension. 
Using Zaatari as a basecamp, our sampled population totals 58,860 (resp. 241,957) in 4 (resp. 16) regions. 



\begin{figure}[ht]
    \centering
    \begin{tabular}{cc}
        {\includegraphics[height=3.7cm] 
        {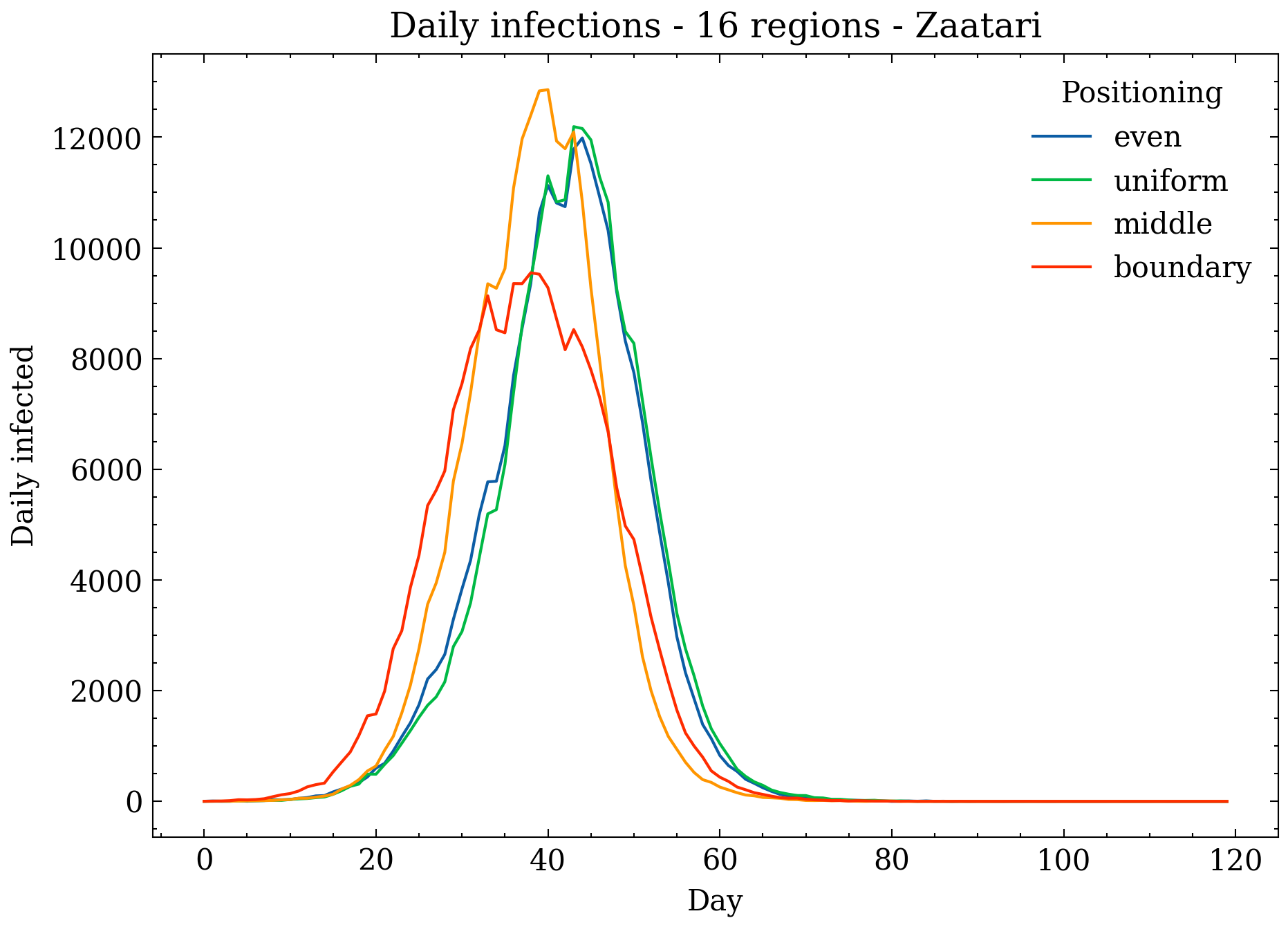}
        }
        &
        {\includegraphics[height=3.7cm] 
        {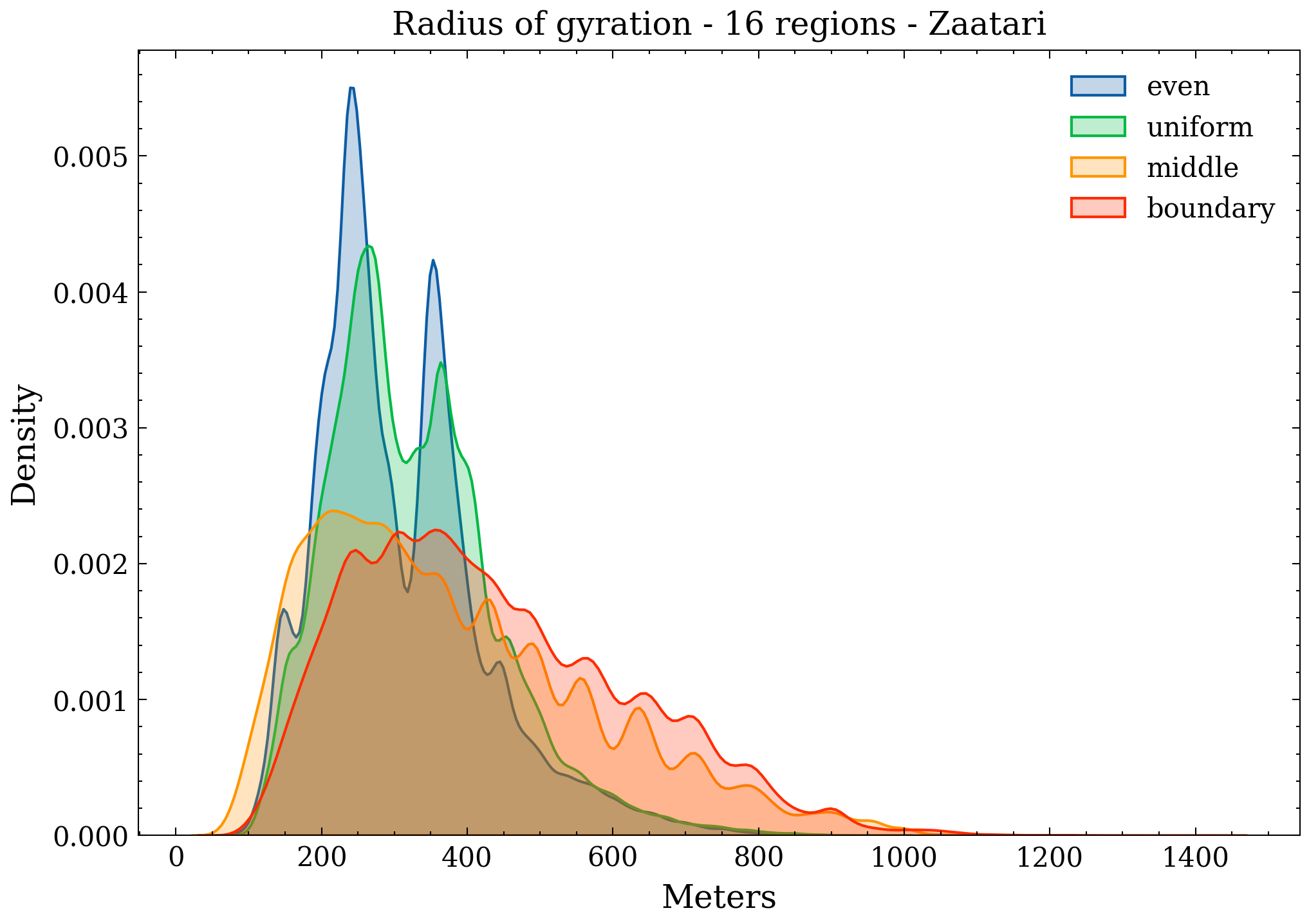}
        }
        \end{tabular}   
    \caption{Infection curves (left) and radius of gyration distribution (right), in 16-regions settlement.}
    \label{fig:zaatari_16_infections_gyration}
\end{figure}

\textbf{Infection curves } 
To gain a first indication of whether camp layout can influence disease spread, we run 120-day long epidemic simulations on virtual camps that only differ by venue positioning schemes, keeping the number of venues fixed as determined by baseline per-capita rates (cf. Table~\ref{tab:per-capita}). 
Infection curves in 4-regions camps are comparable throughout settings, without remarkable differences in terms of height and time of infection peak. 
In contrast, the 16-regions setting (Figure~\ref{fig:zaatari_16_infections_gyration}, left) shows \texttt{boundary} resulting in a substantially lower infection peak than \texttt{middle}. 
Note that \texttt{even} and \texttt{uniform} curves are similar to each other, as one would expect given that both schemes distribute venues throughout the entire camp extension. Appendix~\ref{app:experiments} reports corresponding figures in the 4-regions case (Figure~\ref{fig:zaatari_4_infections_gyration}).
Infection curves disaggregated at sub-regional levels confirm that while disease quickly spreads in a 4-regions camp, in the 16-regions setting the infection travels through the camp and regions peak at different times (Figure~\ref{fig:sub_regional}).

\textbf{Agents' mobility }
Intuitively, visiting nearby venues will have a different epidemiological effect from traveling far across the settlement. 
To quantify this, we measure the reach of our agents in terms of their \emph{radius of gyration}~\citep{gonzalez2008understanding} for different layouts.
We compute this by considering the distances of the facilities assigned to each person by the model, and weighing their contribution by the probability associated with visiting them (see Appendix~\ref{app:gyration} for a detailed formula and an illustrated example). 
In our model, venues assigned to an agent are the 3-nearest (of each type) to the agent's location. 
Figure~\ref{fig:zaatari_16_infections_gyration} (right) shows KDE curves describing the distribution of the radius of gyration in the 16-regions camp. 
As expected, when venues are available more or less regularly throughout the camp interior (\texttt{even} and \texttt{uniform}), people’s average reach is smaller than when facilities are only present at the camp’s middle axes or its boundary. 
Note that in the case of \texttt{middle} and \texttt{boundary} similar gyration patterns do not necessarily translate into similar infection curves.

\begin{wrapfigure}[16]{r}{0.4\textwidth}
\centering
    \includegraphics[height=3.7cm] 
    {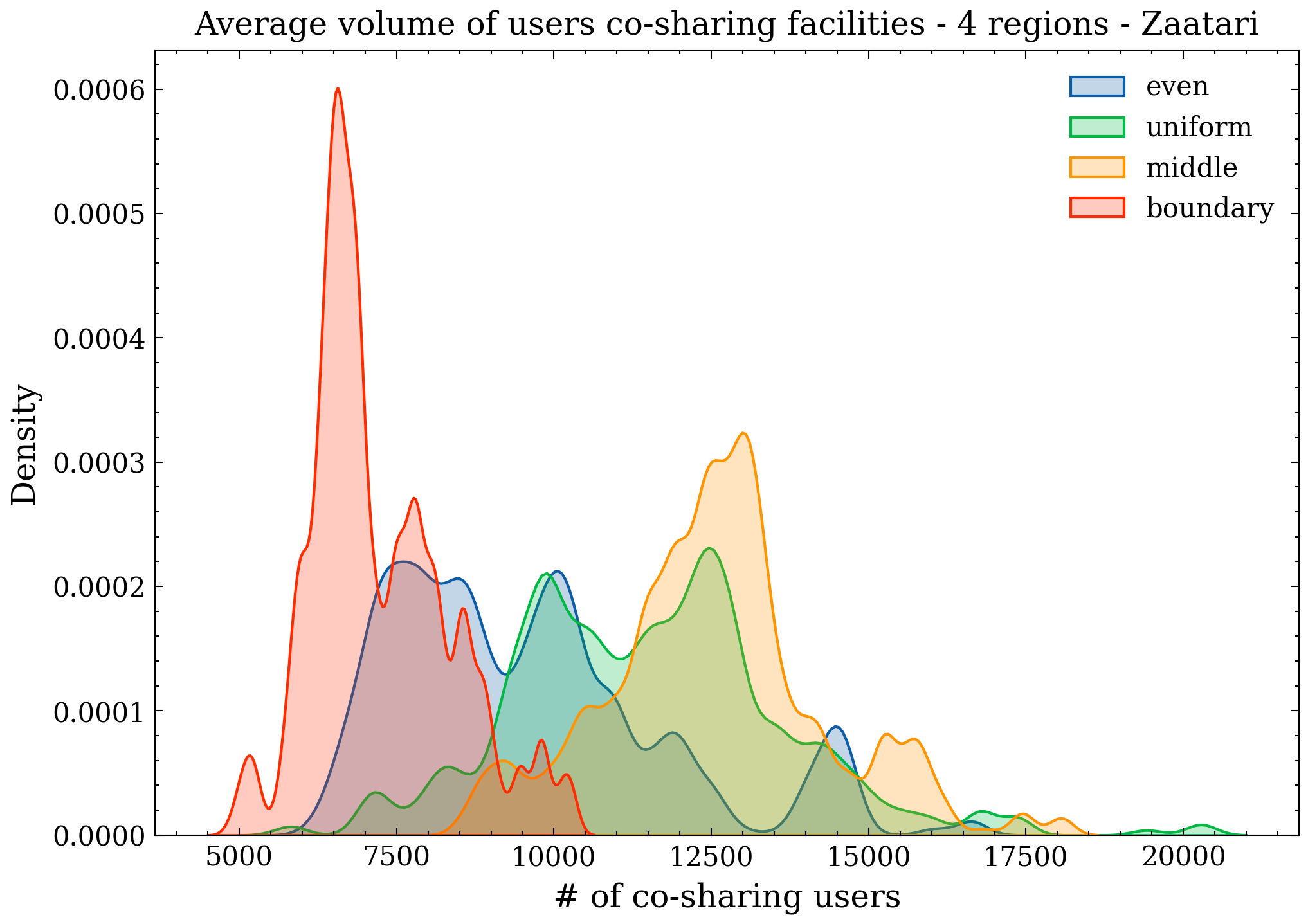}  
\caption{Distribution of the average volume of co-sharing users in the 4-regions setting.}
\label{fig:zaatari_4_sharing}
\end{wrapfigure}

\textbf{Co-sharing users }
We can also interpret agents and venues as nodes of a (bipartite graph) network -- connecting agent $a$ with venue $v$, if $v$ is in the pool of potential venues for $a$ -- to examine how people are connected through their shared venue. 
In the camp setting, some venues (e.g., distribution centers) are necessarily shared by many people, so measuring ``hop-count'' distances between agent nodes is not helpful. 
Instead, by examining the incidence matrix of the camp-defined network, for each person we can compute how many other people (on average) share the same venues they are assigned to. 
We plot KDE curves for this \emph{average number of co-sharing users} in Figures~\ref{fig:zaatari_4_sharing} (and Figure~\ref{fig:zaatari_16_sharing} in Appendix~\ref{app:experiments}). 
We observe a distinct behavior across positioning schemes: \texttt{boundary} induces lower numbers of co-sharing users, while \texttt{middle} results in the volume of users potentially co-sharing (and hence meeting at) shared facilities go up. 
This difference is in line with the infection results observed in Figure~\ref{fig:zaatari_16_infections_gyration} (left), and might contribute to explain why in bigger settlements placing venues along the boundary seems to mitigate disease spread better than placing them in the middle. 




\section{Discussion}
\label{sec:discussion}

This project aimed at creating a methodology to explore the potential effects of different layouts as mitigating factors in the spread of diseases within refugee/IDP settlements. 
Using the \june model, we simulated epidemic spread in virtual camps with facilities distributed according to different prototypical positioning schemes. 
While still at the proof-of-concept stage, the developed framework provides initial recommendations and useful pointers for future investigations.

Intuition supports our finding that camps with limited extensions (4-regions in our experiments) are too small to observe significant differences -- the infection rapidly spreads and dies out within the camp boundaries. 
This is in line with the idea of developing camps with non-contiguous, physically-separated districts\footnote{E.g., as is the case for Azraq Camp (Jordan) \url{https://im.unhcr.org/apps/sitemapping/\#/site/JORs004576/location}}.
However, when settlements grow quickly and spontaneously (the case of Cox’s Bazar), districts’ separation cannot always be guaranteed.
In larger camps (16-regions in our experiments), \texttt{middle} and \texttt{boundary} schemes provide an interesting comparison: while resulting in similar radius of gyration distributions, they give rise to very distinct infection patterns. 
The lower infection curves obtained with \texttt{boundary} might be linked to agents co-sharing their assigned venues with fewer people. 
In real operations, it is conceivable to imagine that a camp originally planned with venues on its boundary could expand in size, and evolve into one in which venues are closer to a \texttt{middle} setup. 
In this sense, re-evaluating this type of assessments in the continuous planning for a camp’s evolution and expansion might play an important role to mitigate disease spread via camp layout design. 



The current framework and experiments lend themselves to be further developed in a number of ways. 
It would be important to validate the present results by running experiments on multiple random seeds, and systematically explore parameter space. 
The region block we introduced could become a modular component to create sites of different shapes and topologies, which would in turn lead to explore other venues’ positioning schemes.
Initial experiments on doubling per-capita rates (e.g., for learning centers) showed that this can also affect infections and agents' usage of the camp. 

The methodology described here is a first step towards a crisis-response tool supporting the design of new settlements with layouts that -- by design -- minimize the impact of infectious diseases. 
To this end, it will be necessary to incorporate the physical environment in which the settlement will be built, as well as additional contextual knowledge. 
For instance, toolkits like~\cite{unhcr-toolkit} 
for security reasons prescribe schools not to be located in peripheral areas, or nearby busy markets and distribution points. 
Ultimately, this type of strategies could also allow to incorporate other concepts of urban planning for the increase of well-being into camps -- as in many occasions people spend a long part of their lives in them. 




\bibliography{iclr2023_conference}
\bibliographystyle{iclr2023_conference}

\newpage
\appendix

\section{Data sources and model assumptions}
\label{app:sources}

\begin{wrapfigure}[14]{l}{0.4\textwidth}
\centering
    \includegraphics[width=\linewidth]{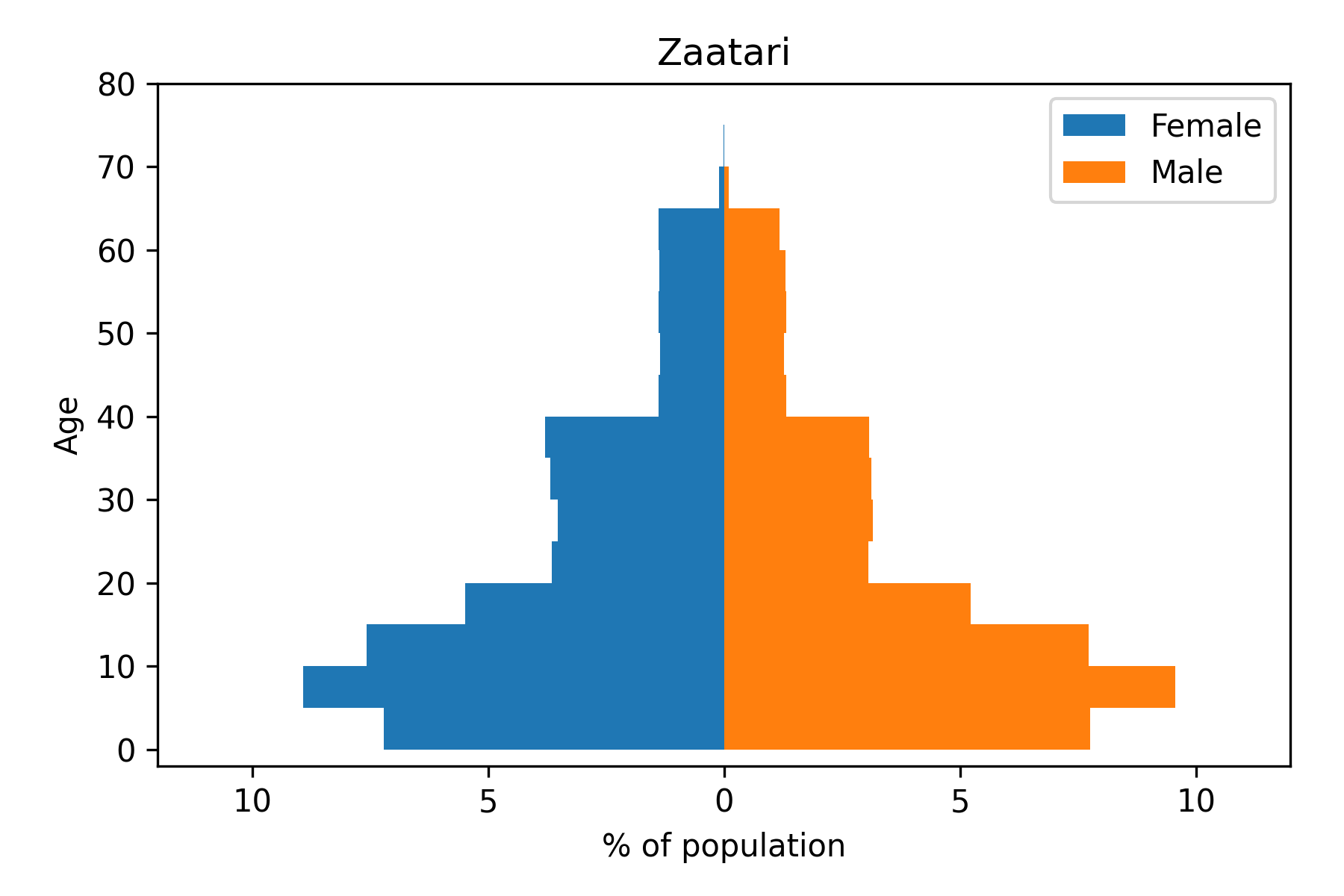}  
\caption{Population age/sex distribution in the Zaatari reference camp.}
\label{fig:zaatari-distribution}
\end{wrapfigure}

\textbf{Demography and household composition }
Statistics on population are based on camps population counts and site assessments, which usually report figures on the camps’ households and inhabitants, disaggregated by age, sex, and their geographic location (in terms of camp blocks and districts). For the Zaatari basecamp, our source is the count conducted by REACH and UNICEF, between December 2014 and January 2015.\footnote{\url{https://data.humdata.org/dataset/za-atari-refugee-camp-population-count}} 
We use a discrete random variable to sample people's age and sex from the basecamp distribution, which is summarized in Figure~\ref{fig:zaatari-distribution}. Further details on the virtual households creation can be found in \citet{walker2022mixed}.

\textbf{About population density }
While we use the basecamp as a guide to determine area-level population counts (fitting a log-normal model to the basecamp distribution of number of area residents), we do not aim to replicate the basecamp population density. 
Instead, we fix the side length of areas to 150 m, so that each region has an extension of 0.81 km\textsuperscript{2} and can accommodate up to 20,000 people while complying with guidelines recommending about 40 m\textsuperscript{2} of open space per person \citep{unhcr-toolkit}. 
As a result, the 4-regions and 16-regions settings we test mainly differ by size of the population and site extension, and not by population density -- which is instead kept in check and compliant with standards. 
The total extension of virtual settlements with 4 (resp. 16) regions is 3.24 km\textsuperscript{2} (resp. 12.96 km\textsuperscript{2}). For reference, the Zaatari camp in Jordan extends for approximately 5.4 km\textsuperscript{2} and is home to almost 81,000 refugees\footnote{\url{https://www.unhcr.org/jo/wp-content/uploads/sites/60/2022/02/1-Zaatari-Fact-Sheet-January-2022-final.pdf}}, while the Cox’s Bazar settlement extends for approximately 13 km\textsuperscript{2} and hosts around 600,000 people in the Kutupalong-Balukhali expansion site\footnote{\url{https://data.unhcr.org/en/documents/details/82872}}.

\textbf{Venue types }
The types of camp venues implemented in our simulation model are listed in Table~\ref{tab:per-capita}, and we refer to \citet{aylett2021operational} for more details. For each facility type, we report the baseline per-capita parameter that we use to determine how many venues of each type should be positioned depending on the total size of the synthesized population. In order to explore scenarios with variable numbers of venues, our framework allows to specify type-specific factors as multipliers to these per-capita rates. 
In our experiments, learning centers are assumed to have 4 shifts throughout the day, with classrooms of 35 pupils maximum. To determine school enrollment rates disaggregated by sex and education level (age) when Zaatari is used as basecamp, we refer to Jordan’s 2020 education indicators as published by the World Bank.\footnote{\url{https://data.humdata.org/dataset/world-bank-education-indicators-for-jordan}}


\textbf{Infection seeding }
In the model, we start the infection by seeding 2 people every day, over the course of 10 days. Instead of uniformly spreading these initial cases, we opt for a clustered approach: we concentrate the seeding in selected households and locate them within a particular region or super-area of the settlement. This scheme aims to
mirror the probable scenario where instead of infections starting uniformly throughout the camp, initial cases might be located within a particular camp district (e.g., maybe one neighboring the host community, or at the boundary of the settlement), with members of some households becoming infected at once. Depending on the extent of the camp, we either select a corner super-area (in the 4-regions case) or a corner region of the settlement (in the 16-regions case).

\textbf{Co-morbidities }
As done in \citet{aylett2021operational,walker2022mixed}, we distribute co-morbidities to the population following the approach outlined in \citet{clark2020global}. For each basecamp, we use data\footnote{\url{https://cmmid.github.io/topics/covid19/Global_risk_factors.html}} relative to the primary country of origin of their refugees/IDP, selecting Syria for Zaatari. 

\textbf{Interactions } 
The contact patterns simulated by the epidemic model are informed by contact matrices that encode how interactions occur in different settings and for different age groups. The contact matrices we use are originally informed by a population survey in the Cox’s Bazar refugee settlement, and then refined with \june’s agent-based model. Their derivation is the focus of \citet{walker2022mixed}.

\textbf{Policy } 
As mentioned in Section~\ref{sec:methodology}, we do not test specific epidemic interventions like social distancing, and we do not assume containment strategies are in place when managing the virtual camp. The only assumptions we make in the model are that (i) hospitalization is available, and (ii) a severely ill person would stay in their shelter and not participate in other camps activities.

\begin{table}[t]
\caption{List of considered venue types with their baseline per-capita parameters and reference source. Note that the per-capita for learning centers and play groups is applied to the children population only (ages between 3 and 17 years, included).}
\label{tab:per-capita}
\begin{center}
\begin{tabular}{llp{0.4\linewidth}}
\toprule
\textbf{Venue type} & \textbf{Baseline per-capita} & \textbf{Reference} \\
\midrule
Distribution center             & 1 / 20,000    & \cite{unhcr-toolkit} \\
Non-food distribution center    & 1 / 20,000    & \cite{unhcr-toolkit} \\
E-voucher outlet                & 1 / 20,000    & Analogous to distribution points  \\
Communal center                 & 1 / 5,000     & \cite{aylett2021operational} \\
Safe space for women and girls  & 1 / 5,000     & \cite{aylett2021operational} \\
Religious center                & 1 / 500       & \cite{aylett2021operational} \\
Learning center                 & 1 / 500       & Derived from \cite{aylett2021operational} and \cite{unhcr-toolkit}\\
Hand-pump and latrine           & 1 / 50        & \cite{unhcr-toolkit} \\
Play group                      & 1 / 20        & \cite{aylett2021operational} \\
Hospital                        & 1 / 20,000    & \cite{unhcr-toolkit} \\
\bottomrule
\end{tabular}
\end{center}
\end{table}

\begin{figure}[t]
\begin{center}
\includegraphics[width=.7\textwidth]{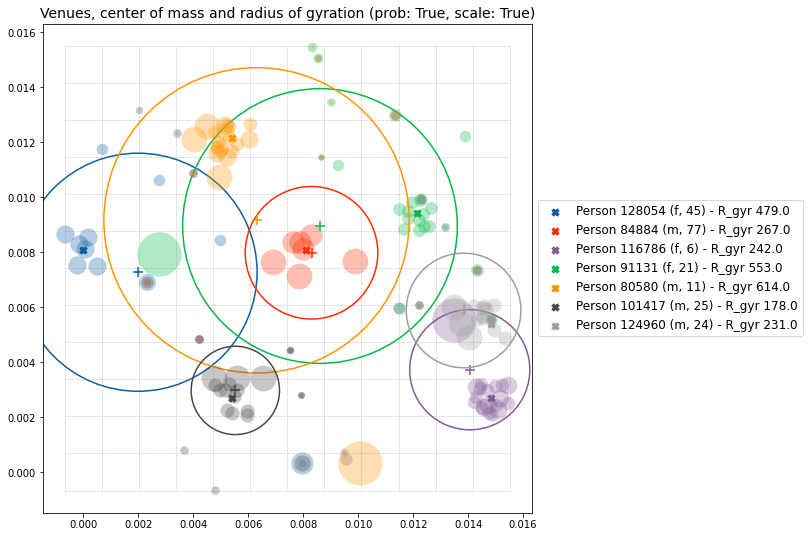}
\end{center}
\caption{Visualizing user mobility with the radius of gyration. For seven agents located in different areas (x markers), assigned venues are identified with colored bubbles of size proportional to their visiting probability. Center of mass (+) and radius of gyration quantify the mobility extent of each user, for which we additionally report sex and age. For example, the average reach of person 101417 (black) is much smaller than that of person 80580 (yellow), the latter being a child assigned to a learning center far away from their household’s area. The underlying grid delineates areas in a 4-regions camp.}\label{fig:gyration-example}
\end{figure}

\section{Radius of gyration}
\label{app:gyration}

To compare people’s mobility in different camp layouts we compute the distribution of the radius of gyration across the population \citep{gonzalez2008understanding}. Measuring the radius of gyration of an agent allows us to characterize its spatial reach within the camp: we take into account the distances of the facilities assigned to each person by the model, and weight their contribution by the probability of visiting them, which is predetermined in the model and depends on the person’s age and sex. 

For a person $u$, let $V_u$ be the set of venues assigned to them by the model. The radius of gyration is computed as 
\begin{equation}\label{eq:gyration}
    rg(u) = \sqrt{\frac{1}{P}\sum_{i\in V_u} p_i \cdot \text{dist}(r_i, r_{cm}(u))^2},
\end{equation}
where for every venue $i\in V_u$ we calculate the distance between its position $r_i$ and the center of mass $r_{cm}$ for the venues assigned to person $u$, and $P = \sum_{i\in V_u} p_i$. The center of mass is given by 
\begin{equation}\label{eq:centermass}
    r_{cm}(u) = \frac{\sum_{i\in V_u} p_i r_i}{\sum_{i\in V_u} p_i},
\end{equation}
using probabilities $p_i$ of going to venue $i$. These probabilities are obtained from the Poisson parameters specified by the \june model, and also act as weights in the radius formula. We equally distribute the probability of going to a venue of a specific type among all those of the same type available in $V_u$. Intuitively, the radius of gyration is small when a user utilizes locations that are close to each other, and gets bigger otherwise. Figure~\ref{fig:gyration-example} provides an illustration of the radius of gyration and its interpretation for seven people in a virtual camp.

We include in the computation of the radius of gyration all venue types in Table~\ref{tab:per-capita}, except hospitals. We include learning centers in the gyration analysis, but because the epidemic model does not define a Poisson parameter for them, we set an analogous parameter with value 1 for students and 4 for teachers, to account for the fact that teachers spend 4 daily shifts (time-steps) at the learning center, while students only have one.

\section{Additional experiments}
\label{app:experiments}

\begin{figure}[hb]
    \centering
    \begin{tabular}{cc}
        {\includegraphics[width=0.4\linewidth]{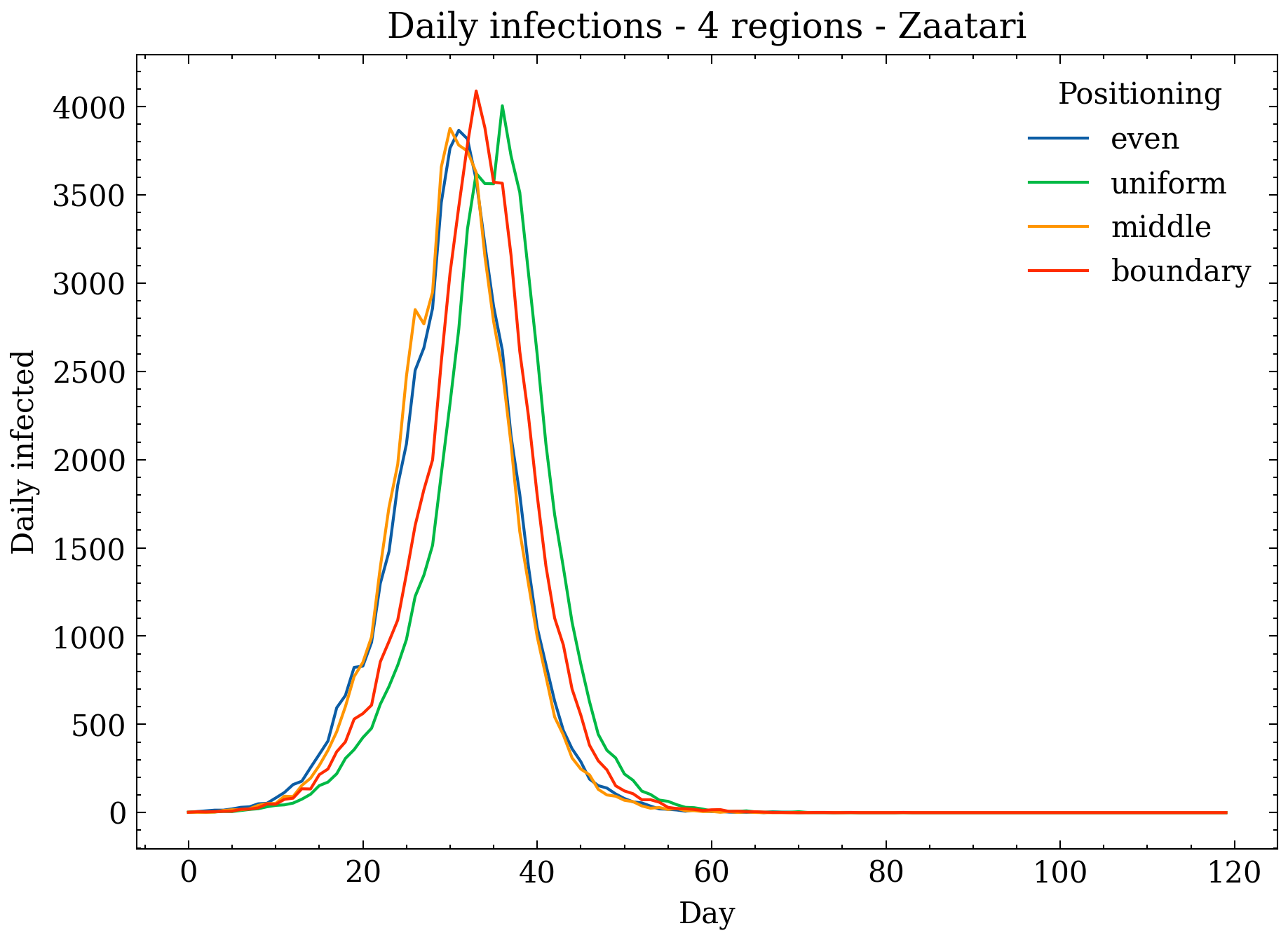}
        }
        &
        {\includegraphics[width=0.4\linewidth]{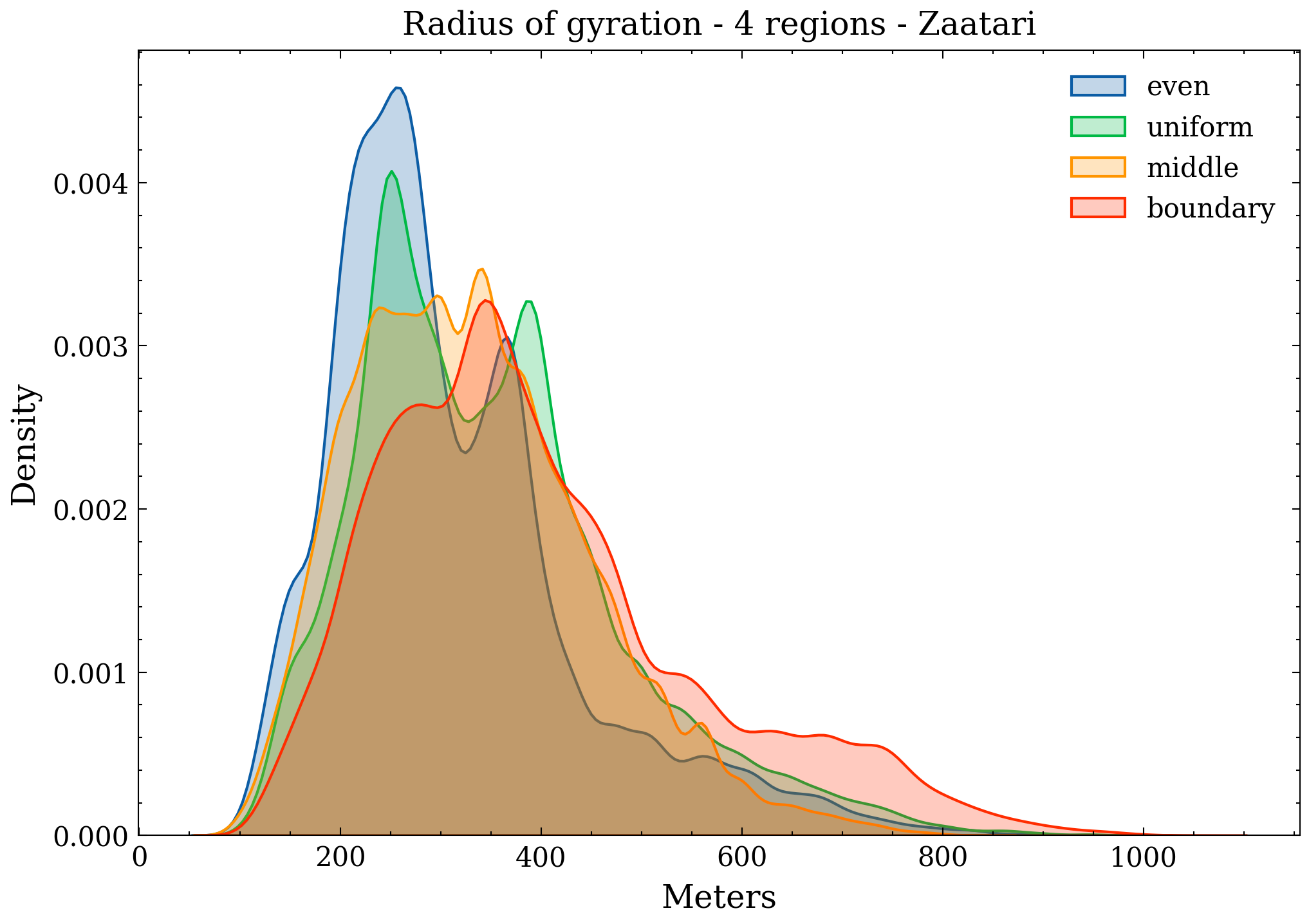}
        }
        \end{tabular}   
    \caption{Infection curves (left) and radius of gyration distribution (right), in 4-regions settlement.}
    \label{fig:zaatari_4_infections_gyration}
\end{figure}

\begin{figure}[t]
\begin{center}
    \begin{subfigure}[b]{0.85\textwidth}
         \centering
         \includegraphics[width=.85\textwidth]{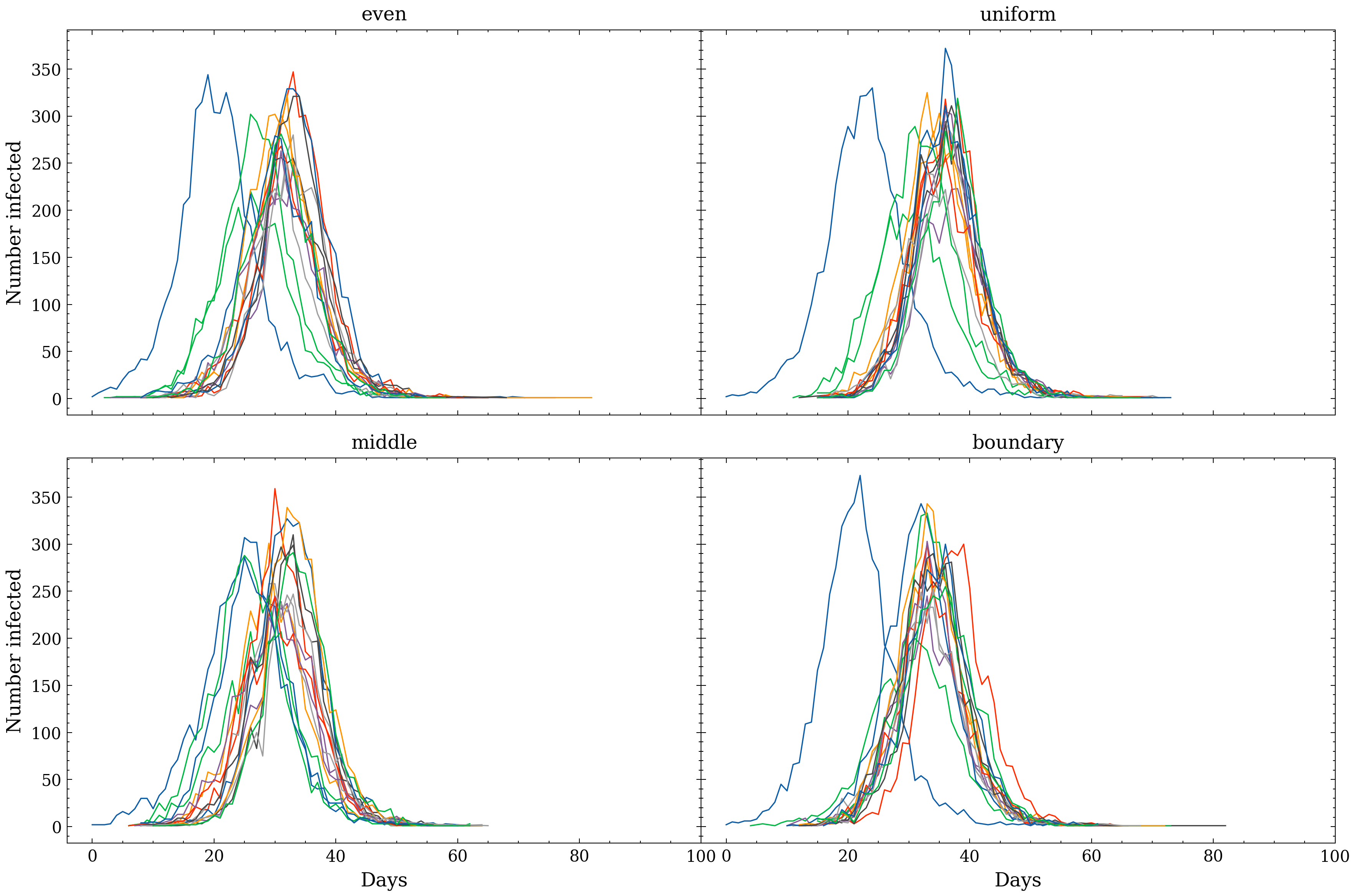}
         \caption{Curves for super-areas in the 4-regions setting.}
         \label{fig:4reg_superareas}
    \end{subfigure}
    \\
    \begin{subfigure}[b]{0.85\textwidth}
         \centering
         \includegraphics[width=.85\textwidth]{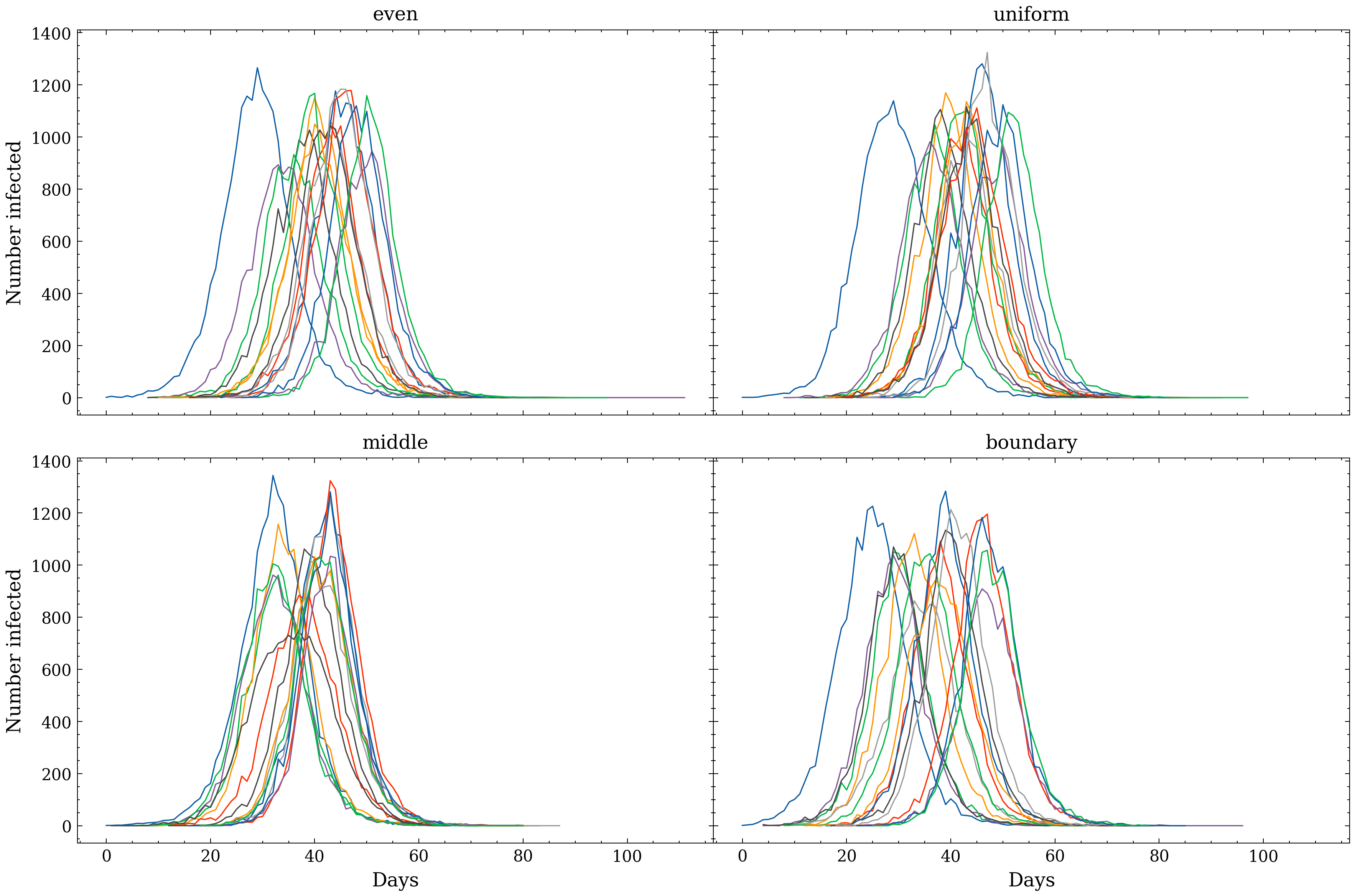}
         \caption{Curves for regions in the 16-regions setting.}
         \label{fig:16reg_regions}
     \end{subfigure}
\end{center}
\caption{Infection curves disaggregated at sub-regional levels.}\label{fig:sub_regional}
\end{figure}

\begin{figure}[ht]
\begin{center}
    \includegraphics[height=4cm]{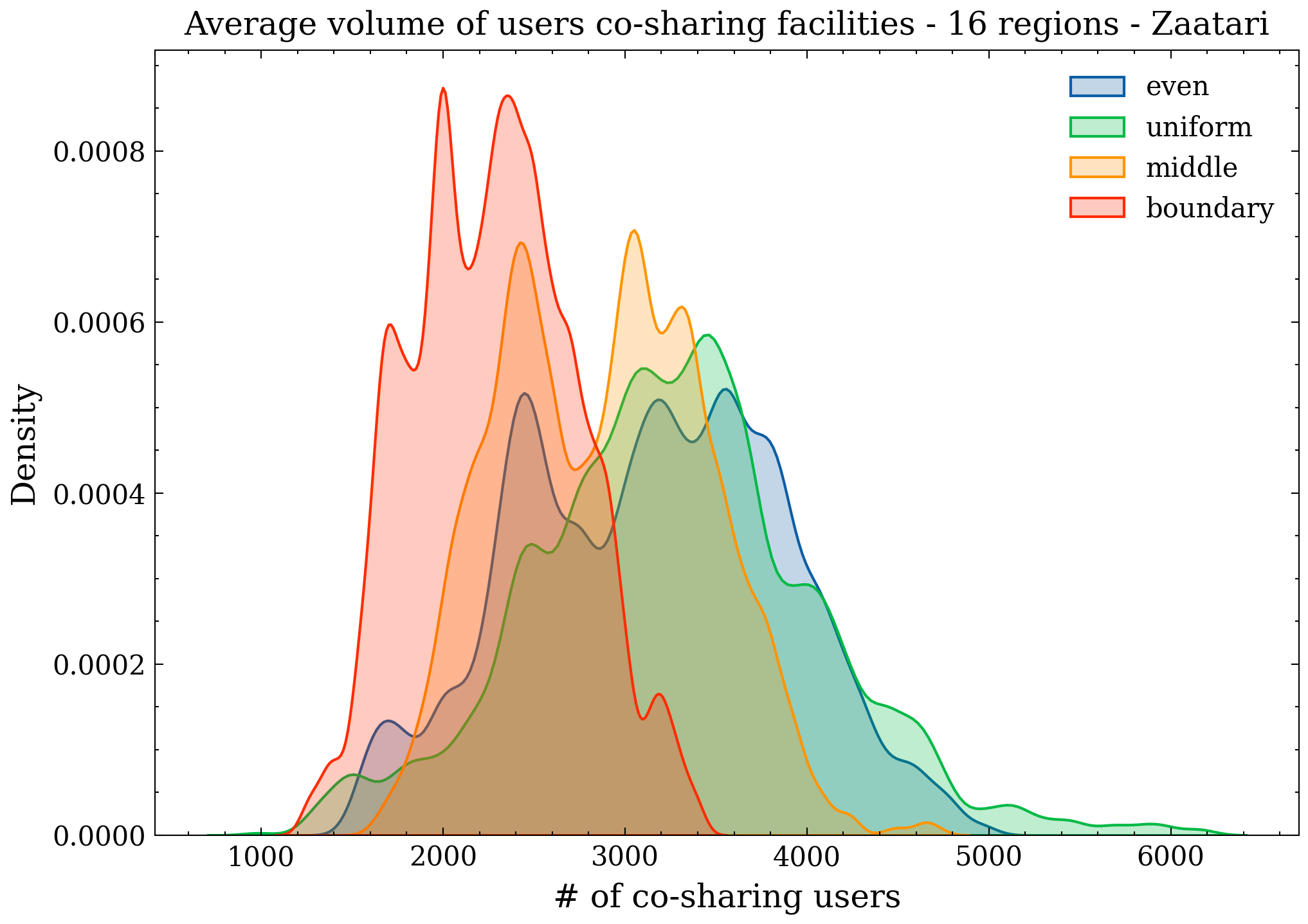}  
\end{center}
\caption{Distribution of the average volume of co-sharing users for a 25\% sample of the population in the 16-regions setting.}
\label{fig:zaatari_16_sharing}
\end{figure}

We collect here additional figures for experiments in both the 4-regions and 16-regions settlements. Figure~\ref{fig:zaatari_4_infections_gyration} is analogous to Figure~\ref{fig:zaatari_16_infections_gyration}, and shows infection curves and radius of gyration distribution in the 4-regions setting. Figure~\ref{fig:sub_regional} reports infection curves disaggregated at sub-regional levels. Figure~\ref{fig:zaatari_16_sharing} is analogous to Figure~\ref{fig:zaatari_4_sharing} and illustrates the distribution of the average number of co-sharing users in the 16-regions setting. Note that for the computation to be tractable in this case, we build an incidence matrix by considering a 25\% sample of the population. 

\end{document}

%% file: math_commands.tex
\usepackage{amsmath,amsfonts,bm}

\def\eqref#1{equation~\ref{#1}}

\def\1{\bm{1}}

\DeclareMathAlphabet{\mathsfit}{\encodingdefault}{\sfdefault}{m}{sl}
\SetMathAlphabet{\mathsfit}{bold}{\encodingdefault}{\sfdefault}{bx}{n}

